\documentclass[10pt,aps,prl,twocolumn,showkeys]{revtex4-1}
\usepackage{amsmath}
\usepackage{graphicx}
\usepackage{epstopdf}
\usepackage{threeparttable}
\usepackage{color}
\usepackage{dcolumn}% needed for some tables
\usepackage{bm}% for math
\usepackage{amssymb}% for math
\usepackage[mediumspace,mediumqspace,Grey,squaren]{SIunits}
\usepackage[normalem]{ulem}

\definecolor{darkerred}{rgb}{0.8,0,0}

\begin{document}

\title{Efficient phonon cascades in hot photoluminescence of WSe$_2$ monolayers}

\author{Ioannis Paradisanos$^{1,2}$}
\author{Gang Wang$^{2,3}$}
\author{Evgeny M. Alexeev$^{2}$}
\author{Alisson R. Cadore$^{2}$}
\author{Xavier Marie$^1$}
\author{Andrea C. Ferrari$^{2}$}
\author{Mikhail M. Glazov$^5$}
%\email{glazov@coherent.ioffe.ru}
\author{Bernhard Urbaszek$^1$}
%\email{urbaszek@insa-toulouse.fr}

\affiliation{%
$^1$Universit\'e de Toulouse, INSA-CNRS-UPS, LPCNO, 135 Avenue Rangueil, 31077 Toulouse, France}
\affiliation{$^2$Cambridge Graphene Centre, University of Cambridge, Cambridge CB3 0FA, U.K.}
\affiliation{$^3$Key Lab of Advanced Optoelectronic Quantum Architecture and Measurement (MOE), School of Physics, Beijing Institute of Technology, Beijing 100081, China}
\affiliation{$^5$Ioffe Institute, 194021, St.-Petersburg, Russia}

\begin{abstract}
Energy relaxation of photo-excited charge carriers is of significant fundamental interest and crucial for the performance of monolayer (1L) transition metal dichaclogenides (TMDs) in optoelectronics. We measure light scattering and emission in 1L-WSe$_2$ close to the laser excitation energy (down to~$\sim$0.6meV). We detect a series of periodic maxima in the hot photoluminescence intensity, stemming from energy states higher than the A-exciton state, in addition to sharp, non-periodic Raman lines related to the phonon modes. We find a period $\sim$15meV for peaks both below (Stokes) and above (anti-Stokes) the laser excitation energy. We detect 7 maxima from 78K to room temperature in the Stokes signal and 5 in the anti-Stokes, of increasing intensity with temperature. We assign these to phonon cascades, whereby carriers undergo phonon-induced transitions between real states in the free-carrier gap with a probability of radiative recombination at each step. We infer that intermediate states in the conduction band at the $\Lambda$-valley of the Brillouin zone participate in the cascade process of 1L-WSe$_2$. The observations explain the primary stages of carrier relaxation, not accessible so far in time-resolved experiments. This is important for optoelectronic applications, such as photodetectors and lasers, because these determine the recovery rate and, as a consequence, the devices' speed and efficiency.
\end{abstract}
\maketitle
\section{Introduction}
Following optical excitation of a semiconductor above the band gap, the subsequent energy relaxation pathways play an important role in optics\cite{klimov1999electron,brida2013ultrafast,kozawa2014photocarrier} and charge carrier transport\cite{song2013transport,glazov2020quantum}. These processes are related to hot charge carriers and excitons and are responsible for the determination of the electron mobility\cite{balkan1998hot}, optical absorption in indirect band gap semiconductors\cite{cardona2010fundamentals}, and intervalley scattering of hot electrons\cite{cardona2010fundamentals}. Photoluminescence (PL) and Raman scattering can be used to probe the interactions of carriers with phonons. In most materials, different types of phonons with different energies can participate in the relaxation process of excited carriers. However, in some materials one type of phonon plays a dominant role and leads to high order processes, e.g., up to 9 longitudinal optical (LO) phonon replicas were reported in the hot PL of CdS and CdSe\cite{leite1969multiple,klein1969multiple,gross1973hot}. Multi-phonon processes are important in defining the optoelectronic performance of ZnO\cite{cerqueira2011resonant,ursaki2004multiphonon,kumar2006photoluminescence, vincent2008raman} and GaN\cite{sun2002outgoing}. Similar effects were measured at 4.2K in bulk MoS$_2$\cite{golasa2014multiphonon}. Reference~\cite{brem2018exciton} predicted that phonon-induced cascade-like relaxation of excitons could be measured in pump-probe experiments of TMDs. Reference~\cite{martin1971cascade} presented a model whereby electrons (holes), e (h), make successive transitions between real states assisted by the emission of a prominent phonon, while other inelastic scattering processes have negligible probability. This gives rise to multiple Stokes-shifted lines.

Group VI transition metal dichalcogenide monolayers (1L-TMDs) are promising for (opto)electronic devices\cite{ferrari2015science,koppens2014photodetectors} due to their direct band gaps in the visible to near-infrared\cite{Splendiani:2010a,Mak:2010a,tonndorf2013photoluminescence}, offering a wide selection of light emission wavelengths at room temperature (RT)\cite{Mak:2010a,lien2019electrical}. Their optical properties are dominated by excitons with binding energies of hundreds of meV\cite{wang2018colloquium}, with spin and valley properties (such as valley-selective circular dichroism\cite{Cao:2012a}) highly beneficial for optoelectronics, valleytronics and spintronics\cite{Novoselov:2016a,Mak:2016a,Schaibley:2016a,unuchek2018room,Schneider2018a,Koperski:2017a,dufferwiel2017valley,Scuri:2018a,hong2014ultrafast,barbone2018charge}.
In 1L-TMDs charge carriers and excitons interact strongly with phonons\cite{song2013transport,he2020valleyphonon,zhang2015chiral,zhu2018observation,trovatello2019strongly}. The optical oscillator strength, i.e. the probability of optical transitions between valence and conduction states\cite{cardona2010fundamentals}, is higher than in III-V quantum wells\cite{cardona2010fundamentals}, resulting in short ($\sim$1ps\cite{Robert:2016a}) exciton lifetimes. This also favors hot PL emission, as excitons relax between several real states\cite{Manca:2017a,han2018exciton}.
\begin{figure*}
\centerline{\includegraphics[width=180mm]{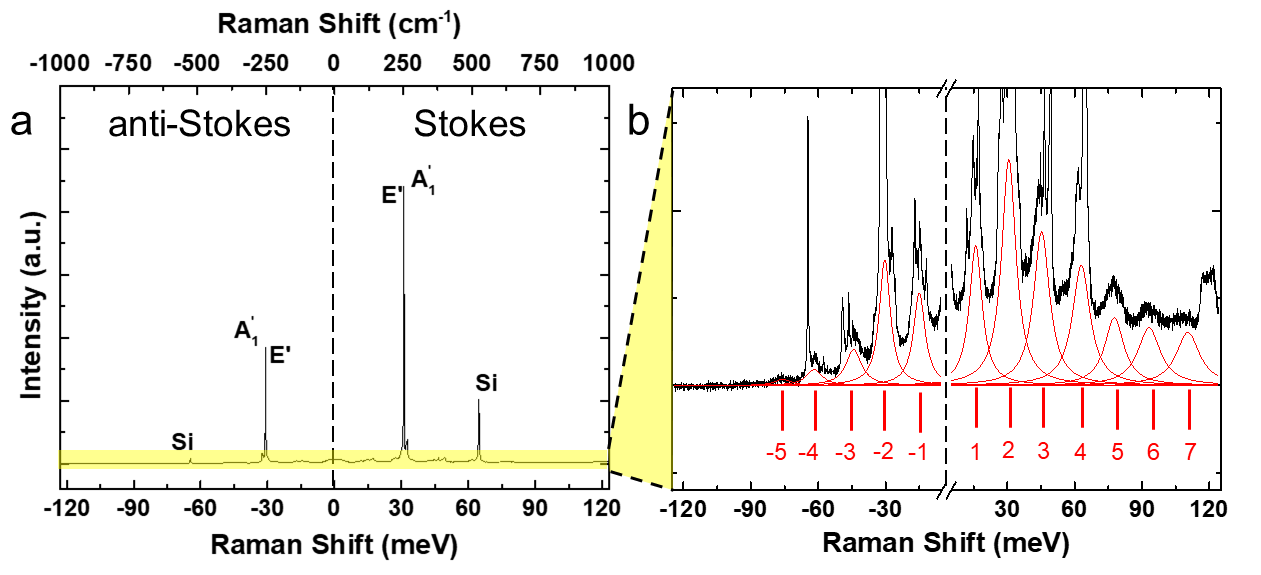}}
\caption{\label{fig:fig1}\textbf{Raman and hot PL spectra of 1L-WSe$_2$ on SiO$_2$/Si} (a) Emission and scattering spectrum of 1L-WSe$_2$ at 295K as a function of energy shift with respect to the excitation laser (2.33eV). The degenerate in-plane ($E'$) and out-of-plane ($A_1'$) Raman mode$\sim250cm^{-1}$\cite{tonndorf:2013}, as well as the Si Raman peak$\sim521cm^{-1}$\cite{temple1973multiphonon}, are prominent in both S and AS. (b) Magnified portion of the spectrum in yellow in (a). This reveals 7 periodic S peaks and 5 AS. Their intensity decreases as a function of the energy shift for both S and AS}.
\end{figure*}
\begin{figure*}
\centerline{\includegraphics[width=180mm]{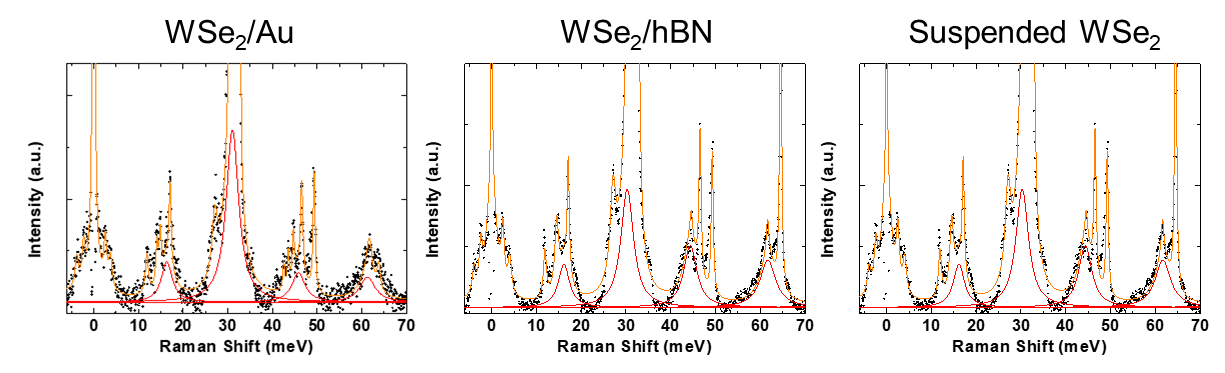}}
\caption{\label{fig:figS3} \textbf{Raman spectra of 1L-WSe$_2$ on different substrates}, at 295K and 514nm. Black points are the experimental data. The red lines are the fitted cascades and yellow line is the sum of the fitted Lorentzians}
\end{figure*}
\begin{figure}
\centerline{\includegraphics[width=90mm]{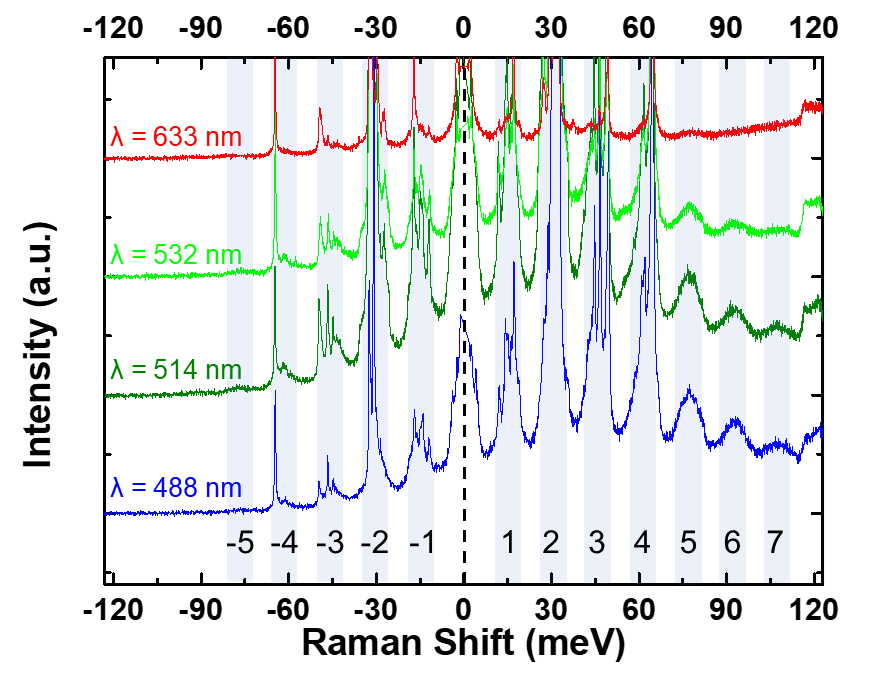}}
\caption{\label{fig:figS4} \textbf{Phonon cascades for different laser energies.} Raman spectra of 1L-WSe$_2$ on SiO$_2$/Si at 488, 514, 532, 633nm and 295K, shifted vertically for clarity}
\end{figure}
\begin{figure*}
\centerline{\includegraphics[width=180mm]{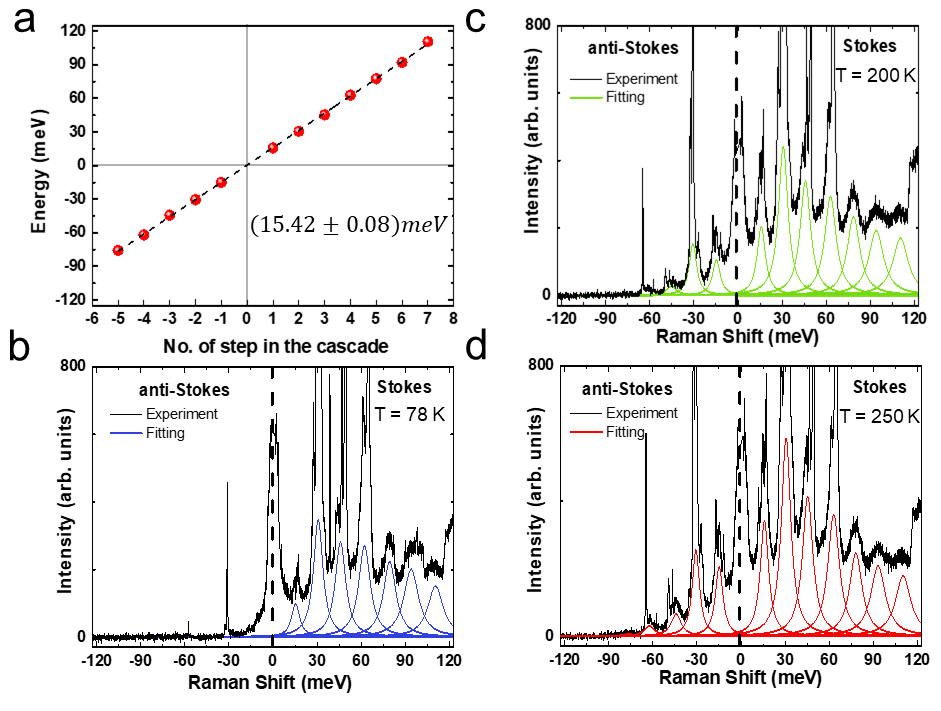}}
\caption{\label{fig:fig2}\textbf{Energy separation and T dependence} (a) Emission energies as a function of number of steps in the cascade, extracted from the RT spectrum in Fig.\ref{fig:fig1}b. The dashed black line is a linear fit, giving a step energy$\sim$15.42$\pm$0.08meV.  (b-d)Hot PL spectra of 1L-WSe$_2$ at (b) 78K, (c) 200K, (d) 250K for 532nm excitation}
\end{figure*}

Here, we use an ultra-low ($\sim$5cm$^{-1}\sim$0.6meV) cut-off frequency (ULF) Raman spectroscopy system (Horiba LabRam HR Evolution) to investigate the light scattered and emitted by 1L-WSe$_2$ on SiO$_2$, hBN and Au, as well as suspended 1L-WSe$_2$. We observe phonon-assisted emission of hot PL periodic in energy both in the Stokes (S) and anti-Stokes (AS) spectral range, and we extract a phonon energy $\sim$15meV. The S signal shows 7 maxima at 78$\ldots$295K. We also detect up to 5 maxima in the anti-Stokes signal $\sim$75meV above the laser excitation energy. The AS signal increases in intensity as the temperature (T) is raised. In order to explain the findings, we extend the theory of the cascade model initially developed for bulk crystals\cite{martin1971cascade}, to 1L-TMDs. We include finite T effects to compare S and AS signals and to understand carrier relaxation at RT. By analyzing the T and excitation energy dependence of our spectra, we conclude that a continuum of states (in the free-carrier gap) is involved in the e-h relaxation in 1L-WSe$_2$. Intermediate states in the conduction band around the $\Lambda$-valley of the Brillouin zone (BZ) participate in the cascade process. Hot PL so close in energy to the non-resonant excitation laser gives access to the initial stages of carrier relaxation. These processes are normally ultrafast (e.g. $\sim$100fs in GaAs\cite{kash1985subpicosecond}) and challenging to be traced by time-resolved experiments. Understanding the carrier relaxation pathways in 1L-WSe$_2$ is important for optoelectronic applications, such as photodetectors\cite{koppens2014photodetectors} and lasers\cite{reeves20182d}, because it determines the recovery rate (i.e. the population of carriers relaxing to the ground state over time) and, as a result, the devices' speed and efficiency.
\section{Results and Discussion}
1L-WSe$_2$ flakes are exfoliated from bulk 2H-WSe$_2$ crystals (2D Semiconductors) by micromechanical cleavage on Nitto Denko tape\cite{novoselovPNAS2005}, then exfoliated again on a polydimethylsiloxane (PDMS) stamp placed on a glass slide for inspection under optical microscope. Optical contrast is used to identify 1L prior to transfer\cite{casiraghi2007rayleigh}. Before transfer, 85nm (for optimum contrast\cite{casiraghi2007rayleigh}) SiO$_2$/Si substrates are wet cleaned\cite{purdie2018cleaning} (60s ultrasonication in acetone and isopropanol) and subsequently exposed to oxygen-assisted plasma at 10W for 60s. The 1L-WSe$_2$ flakes are then stamped on the substrate with a micro-manipulator at 40$^{\circ}$C, before increasing T up to 60$^{\circ}$C to release 1L-WSe$_2$\cite{orchin2019niobium}. The same procedure is followed for transfer of 1L-WSe$_2$ on hBN, Au and Si substrates with 2$\mu$m Au trenches made by lithography, to suspend the samples.
\begin{figure*}
\centerline{\includegraphics[width=180mm]{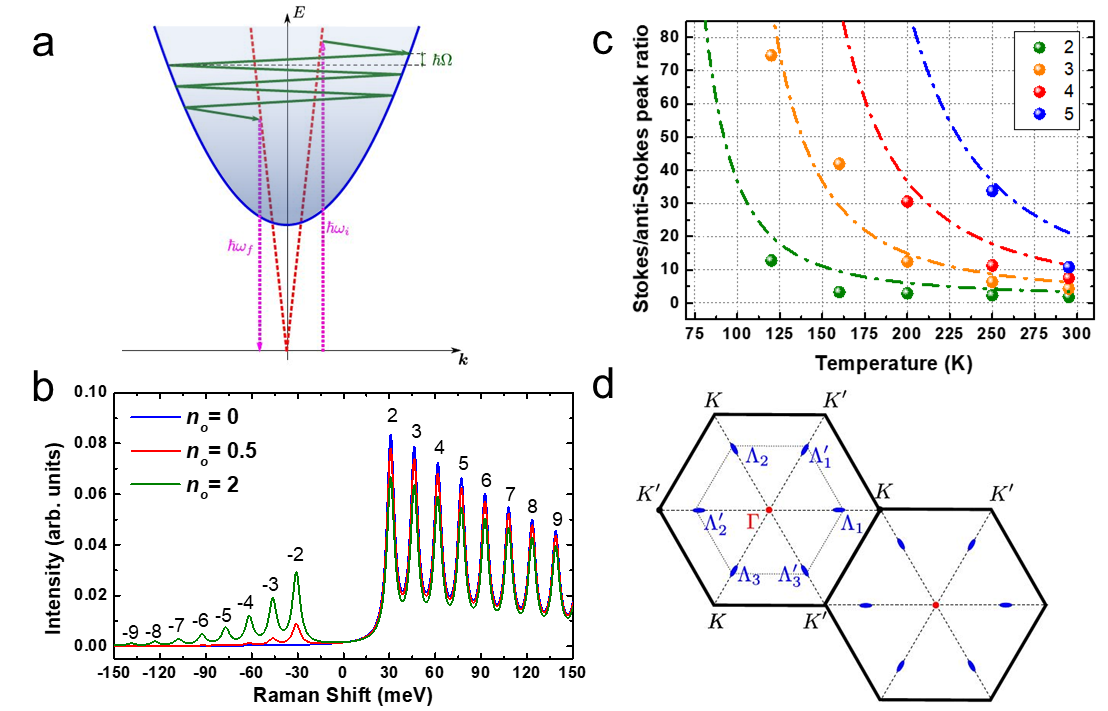}}
\caption{\label{fig:fig3} \textbf{Comparison between experiments and theory} (a) Scheme of phonon-assisted hot PL. The incident, $\hbar\omega_i$, and outgoing, $\hbar\omega_f$, photons are shown by dotted magenta vertical arrows. The phonons participating in the cascade are indicated by the green arrows.  The e-h pair dispersion curve is the blue parabola. The light cone is shown by red dashed lines. (b) Calculated S/AS spectrum at different T. (c) I$_S$/I$_{AS}$ for different numbers of cascade steps as a function of T. Filled circles are experimental data at 532nm. The fit with Eq.\eqref{ratio} is indicated by dot-dashed lines. (d) Extended BZ of 1L-WSe$_2$. Corresponding valleys are marked as $\Gamma$, $K$, $K'$ and $\Lambda_i$, $\Lambda_i'$ ($i=1,\ldots,3$)}
\end{figure*}

The Raman and hot PL spectra are recorded in a back-reflection geometry with a 50X objective (NA=0.45) and a spot size$\sim1\mu$m. A liquid nitrogen cryostat (Linkam Scientific) placed on a XY translational stage is used to control $T=78\ldots295$K and excitation area. Imaging of the sample and monitoring of the excitation spot position are achieved using a set of beam splitters, aligned to a charge-coupled device (CCD) camera. The PL and Raman signals collected in the backward direction are filtered by 3 notch volume Bragg filters with a total optical density (OD)=9. The cut-off frequency is $\sim5$cm$^{-1}\sim 0.6$meV. The filtered signals are then focused on the slit of the spectrometer and dispersed by a 1800l/mm grating before being collected by the detector.

A typical RT Raman spectrum for 1L-WSe$_2$ on SiO$_2$/Si measured at 532nm is in Fig.\ref{fig:fig1}a. The degenerate in-plane, $E'$, and out-of-plane, $A_1'$, modes of 1L-WSe$_2$\cite{tonndorf:2013} dominate the spectrum$\sim250$cm$^{-1}$ ($-31$meV) and$\sim250$cm$^{-1}$ ($+31$meV) in the AS and S range. Rescaling the intensity within the region marked in yellow in Fig.\ref{fig:fig1}a reveals an underlying periodic pattern, Fig.\ref{fig:fig1}b. Hereafter, for the energy scale we use meV instead of cm$^{-1}$.

We fit all the peaks between $-120$meV and $+120$meV using Lorentzians (see Methods), as shown in red in Fig.\ref{fig:fig1}b. There are 7 S peaks and 5 AS at 295K. The peak$\sim$120meV ($\sim$970cm$^{-1}$) originates from a combination of the Si substrate $ \Gamma_{1},\Gamma_{12},\Gamma_{25'} $ phonons\cite{temple1973multiphonon}. Although the energy separation between two consecutive peaks is constant, the intensity decreases as a function of energy with respect to the excitation energy (here fixed at 0). To exclude other contributions, such as thin-film interference effects\cite{klar2013raman,Robert:2018a}, we measure 1L-WSe$_2$ transferred on Au, suspended, and placed on few-layer (FL) ($\sim$10nm) hBN, Fig.\ref{fig:figS3}. The intensity of the hot PL is comparable among the same steps of the cascade and the position of the peaks is the same. Therefore, the cascade is linked to intrinsic relaxation mechanisms of 1L-WSe$_2$ and not to substrate-induced interference. Henceforth we will focus on 1L-WSe$_2$ on SiO$_2$/Si.

To exclude the possibility that our laser is in resonance with a specific transition, we perform variable excitation wavelength experiments at 295K. Figure~\ref{fig:figS4} plots the spectra measured at 488nm ($\sim$2.54eV), 514nm ($\sim$2.41eV), 532nm ($\sim$2.33eV) and 633nm ($\sim$1.96eV). We observe the same high-order features with identical energy separations in both S and AS. All these excitation energies\cite{633attention} lie above the free carrier gap of 1L-WSe$_2\sim$1.89eV\cite{goryca2019revealing,Wang:2015b,He:2014a}. Fig.\ref{fig:fig2}a plots the energy offset with respect to the excitation laser (here 532nm) of each emission feature as a function of the number of steps in the cascade at 295K. Applying a linear fit, we extract$\sim$15.42$\pm$0.08meV, regardless of substrate and excitation energy. This strong periodic modulation of the detected light intensity suggests that the scattering of photoexcited carriers is dominated by one prominent phonon mode. Since we excite above the free carrier gap of 1L-WSe$_2$\cite{goryca2019revealing}, the intermediate states of the transitions are real. The e-h pair representation is depicted in Fig.\ref{fig:fig3}a.

The lattice T could affect the peaks intensity, as phonon occupation increases with T\cite{jellison1983importance,kip1990determination}. We thus perform T dependent measurements from 78 to 295K, while keeping the excitation power constant$\sim$26$\mu$W. No emission AS features are observed at 78K, Fig.\ref{fig:fig2}b, with the exception of two sharp lines$\sim-30$ and$\sim-60$meV, originating from 1L-WSe$_2$ and Si Raman modes, respectively. The hot PL peaks are clearly seen at 200K, Fig.\ref{fig:fig2}c, and a further increase in intensity is observed at 250K, Fig.\ref{fig:fig2}d. Additional measurements at 120, 160, and 295K are performed and used in the fits in Fig.5.

At low T (78K), phonon \textit{absorption} processes are suppressed because of the insufficient lattice thermal energy\cite{jellison1983importance}. Optical excitation results in free e-h pair formation\cite{steinleitner2017direct,trovatello2020ultrafast} or virtual formation of an exciton with small in-plane wavevector ($k\lesssim\omega_i/c$ with $\omega_i$ the excitation laser frequency). With the subsequent phonon emission, the e-h pair reaches a real final state (blue parabola in Fig.\ref{fig:fig3}a), for which radiative recombination is forbidden by momentum conservation. This triggers the cascade relaxation process, whereby at each step a phonon is emitted (or absorbed at elevated T). If the interaction with one phonon mode with energy $\hbar\Omega$ dominates over all other inelastic scattering processes, the exciton loses energy by integer multiples of $\hbar\Omega$\cite{martin1971cascade,cardona2010fundamentals}. After emission of several ($\geqslant$2) phonons, the exciton recombines and emits a photon with frequency $\omega_f$ in a two-step process via an intermediate state with a small wavevector, for which radiative recombination is momentum allowed. Thus, we have secondary emission or scattering of light with S shift $\omega_i-\omega_f=j\Omega$, where $j=2,3, \ldots$. $j=\pm1$ is impossible as we scatter out of the light cone (i.e. light linear dispersion) with the first event. At finite T, in addition to phonon emission, phonon absorption also comes into play and AS emission is observed at $\omega_f-\omega_i=j\Omega$.

Multiphonon processes that do not involve real states require higher order in the exciton-phonon interaction\cite{shree2018observation}, and are therefore less probable. In contrast, the process in Fig.\ref{fig:figS4} is resonant, since excitation in the free-carrier gap means all intermediate states are real. This allows us to describe the phonon emission cascade via the kinetic equation for the exciton distribution function $f(\varepsilon)$, where $\varepsilon$ is the exciton energy, as derived in Methods. Since the energy of the exciton changes in each scattering event by $\pm \hbar\Omega$, the distribution function can be written as:
\begin{equation}
\label{distribution}
f(\varepsilon) = \sum_{j=-\infty}^\infty f_j \delta(\varepsilon_0- j\hbar\Omega)
\end{equation}
where $\varepsilon_0$ is the excitation energy, $\delta(\varepsilon)$ is the Dirac $\delta$-distribution (the phonon dispersion and phonon damping results in the broadening of the $\delta$-distribution,as detailed in Methods), $f_j$ describe the peaks intensity. At the steady state (partial derivative with respect to time equals zero) these obey a set of coupled equations describing the interplay of in- and out-scattering processes:
\begin{multline}
\label{fj:eqs0}
\gamma f_j = \gamma_o \left[f_{j-1} (n_o+1) + f_{j+1} n_o\right] + g\delta_{j,0},\\
 \quad j=\ldots, -2,-1,0,1,2,\ldots.
\end{multline}
where $n_o =\left[\exp{\left({\hbar\Omega}/{k_B T}\right)} -1\right]^{-1}$ is the phonon mode occupancy at $T$, $\gamma_o$ is the rate of the spontaneous phonon emission, $\gamma=\gamma_o (2n_o+1)+\gamma'$, is the total damping rate of the exciton, which includes recombination and inelastic scattering processes $\gamma'$. The last term in Eqs.\eqref{fj:eqs0}, $g\delta_{j,0}$, describes the exciton generation at the energy $\varepsilon_0$, and is proportional to the exciton generation rate. Equation~\eqref{fj:eqs0} has the boundary conditions:
\begin{equation}
\label{fj:bound}
\lim_{j\to -\infty} f_j = 0, \quad f_{K+1} =0,
\end{equation}
where $K$ is the maximum number of steps in the cascade:
\begin{equation}
\label{steps:K}
K=\left\lfloor \frac{\hbar\omega_i - E_1}{\hbar\Omega}\right\rfloor,
\end{equation}
with $E_1$ the energy of the exciton band bottom. Equations~\eqref{fj:eqs0} are derived assuming $\gamma_o$ and $\gamma'$ independent of $\varepsilon$. This assumption is needed to get an analytical solution of Eqs.\eqref{fj:eqs0}, but can be relaxed, as discussed in Methods.

The general solution of Eqs.\eqref{fj:eqs0} is:
\begin{equation}
\label{fj:sol}
f_j = \begin{cases}
A x_+^j, \quad j<0,\\
B x_+^j + C x_-^j, \quad j\leqslant 0,\\
\end{cases}
\end{equation}
where
\begin{equation}
\label{xpm}
x_\pm = \frac{\gamma \pm \sqrt{\gamma^2-4n_o(n_o+1)\gamma_o^2}}{2 \gamma_o n_o},
\end{equation}
and $x_+>1$ and $x_-<1$,  $A$, $B$, and $C$ are the coefficients. For cascades with $K\gg1$ we can set $B=0$ and
\begin{multline}
A=C=\frac{g}{\sqrt{\gamma^2-4n_o(1+n_o)\gamma_o^2}}\\
=\frac{g}{\sqrt{\gamma_o^2+\gamma'^2+2\gamma_o\gamma'(1+2n_o)}}.
\end{multline}
In this model, the spectrum of the scattered light consists of peaks with I$\propto f_j$, with scattering cross-section:
\begin{multline}
\label{sigma}
\sigma(\omega_i,\omega_f) = \sigma_0(\omega_i,\omega_f) \\
\times \left.\sum_{j=2}^\infty\right.' \frac{1}{\pi} \frac{2\Gamma}{4\Gamma^2+(j\Omega -\omega_i + \omega_f)^2} f_j.
\end{multline}
Here $\sigma_0(\omega_i,\omega_f)$ is a smooth function of frequency, $\Gamma$ is the phonon damping. This description is valid for peaks with $|j|>1$, the prime at the summation denotes that the terms with $j=0,\pm 1$ are excluded. Accordingly, the peaks with Raman shift $\pm \hbar\Omega$ are suppressed. At $n_o\to 0$ (limit of low T), $x_+\gg 1$ and $\mathrm I_j$ with negative $j$ (AS components) are negligible. At the same time, $x_- \to (\gamma_o/\gamma)$ and the S peak intensities, I$_S$, scale as $(\gamma_o/\gamma)^j$. This scaling is natural for cascade processes\cite{cardona2010fundamentals,ivchenko_lang_pavlov77,Goltsev_1983}, since the probability of phonon emission relative to all other inelastic processes is given by $\gamma_o/\gamma$, thus I$_S$ decays in geometric progression. At finite T, the AS peaks appear with I$_{AS}$ proportional to the thermal occupation of the phonon modes. Thus I$_S$/I$_{AS}$ with $j$ steps in the cascade can be written as:
\begin{equation}
\label{ratio}
\frac{\mathrm I_{S}(j)}{\mathrm I_{AS}(j)}=\frac{f_j}{f_{-j}} = \left(1+\frac{1}{n_o}\right)^j,
\end{equation}
and corresponds to the ratio of phonon emission and absorption rate to the power of $j$.

The calculated I distribution and spectra at various T (corresponding to different $n_o$) are in Fig.\ref{fig:fig3}b. Fig.\ref{fig:fig3}c plots I$_S$/I$_{AS}$ as a function of T from Eq.\eqref{ratio}. The experimental points collected from the fitted I of each step in the cascade at 532nm excitation are displayed with circles. The absence of data at 78K indicates no detection of I$_{AS}$ at this T. Applying Eq.\eqref{ratio} to the steps 2 to 5 in the cascade with a phonon energy$\sim$15.4meV extracted from Fig.\ref{fig:fig2}a, gives the dashed lines in Fig.\ref{fig:fig3}c, in good agreement with experiments.
\begin{figure}
\centerline{\includegraphics[width=90mm]{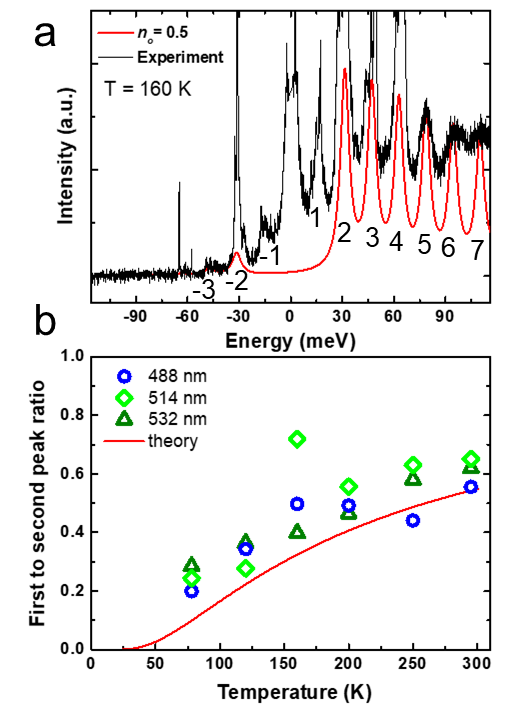}}
\caption{\label{fig:figS8} (a) Experimental data at 160K (black line) compared to the calculated spectrum from Eq.\eqref{fj:eqs0} for $n_o$=0.5. (b) Ratio of measured intensities of $j=1$ to $j=2$ peaks and corresponding fit with Eq.\eqref{rat12}}
\end{figure}

Our model well captures the main experimental observations. The periodic pattern of hot PL I is reproduced by the calculations, Fig.\ref{fig:fig3}b, and I$_S$/I$_{AS}$ closely follows Eq.\eqref{ratio}, Fig.\ref{fig:fig3}c. There is good agreement between our experimental data and the calculated spectra from Eq.\eqref{fj:eqs0}. An example for $n_o$=0.5 at 160K is in Fig.\ref{fig:figS8}a. In our model, the peaks with $j=\pm 1$ are absent because $N\geqslant2$ phonons are needed for the first step of the cascade process, as for Fig.\ref{fig:fig3}a. Fig.\ref{fig:fig1} shows that $j=\pm 1$ peaks are smaller than $j=\pm 2$ ones, but still detectable. One scenario disregarded in our model is a process where the two phonons are emitted and then one is absorbed (or vice versa). In this case $\mathrm I_{j=1}/\mathrm I_{j=2}$ should be strongly T dependent in the $78\ldots295$K range. This is indeed the case in our experiment, Fig.\ref{fig:figS8}b. This additional channel is also based on the interaction with the same phonon energy$\sim$15meV. Another possible effect is elastic scattering of excitons by disorder or acoustic phonons, whereby exciton transitions in and out of the light cone can be controlled by elastic scattering (see Methods for details).

To get a better understanding of the relaxation pathways, we consider different scattering mechanisms.

Scattering within the same valley is not plausible due to the mismatch of BZ centre phonon energies\cite{he2020valleyphonon}. $\sim$15meV could correspond to either $\Gamma-K$ or $\Gamma-\Lambda$ phonons. The phonon dispersion in 1L-WSe$_2$ show acoustic phonons with energies$\sim$15meV\cite{he2020valleyphonon,PhysRevB.90.045422}. These have a flat dispersion, necessary to observe the high number of oscillations we report, and are compatible with the model in Fig.\ref{fig:fig3}a.

Another option involves $K$-$K'$ scattering of e (h) or, equivalently, $\Gamma$-$K$ scattering of excitons. This would result in I oscillations as a function of the step in the cascade, due to the suppression of the process $\Gamma\rightarrow K\rightarrow K' \rightarrow \Gamma$ compared to $\Gamma\rightarrow K\rightarrow\Gamma$ (see Methods for details). However, we do not observe I oscillations for different cascade steps in our spectra. As a result, we exclude this scenario. Therefore, the excitonic states in the $\Lambda$ valleys play a role as intermediate states, Fig.\ref{fig:fig3}d. The conduction band minima in these valleys are relatively close ($\sim$35meV) to $K$ and play a crucial role in exciton formation and relaxation\cite{Kormanyos:2015a,selig2018dark,madeo2020directly,lindlau2017identifying,rosati2020temporal}. In this case, h remain in $K$ (or $K'$), but e scatter to any of the 6 available $\Lambda$ valleys and then scatter between these $\Lambda$ valleys before going back to $K$ ($K'$). This can be described taking into account all pathways, as:
\[
\mbox{photon} \to \Gamma \xrightarrow{\hbar\Omega} \underbrace{\Lambda_i \xrightarrow{\hbar\Omega} \ldots \xrightarrow{\hbar\Omega} \Lambda_j'}_{j} \xrightarrow{\hbar\Omega}  \mbox{photon},
\]
with arbitrary number of steps $j$ (both odd and even). The matrix elements of the processes are similar.

Similar oscillations can appear for free e and h\cite{martin1971cascade}. The basic description of the effect is similar to what we observe here, and our model can be extended to take into account the e/h distribution functions. The spectra of scattered light and I$_S$/I$_{AS}$ are similar to those calculated above. We cannot distinguish between exciton and the free carrier cascades directly in our experiments. The excitonic description, however, seems straightforward due to enhanced (with respect to bulk materials) Coulomb effects in 1L-TMDs\cite{wang2018colloquium}.
\begin{figure*}
\centerline{\includegraphics[width=180mm]{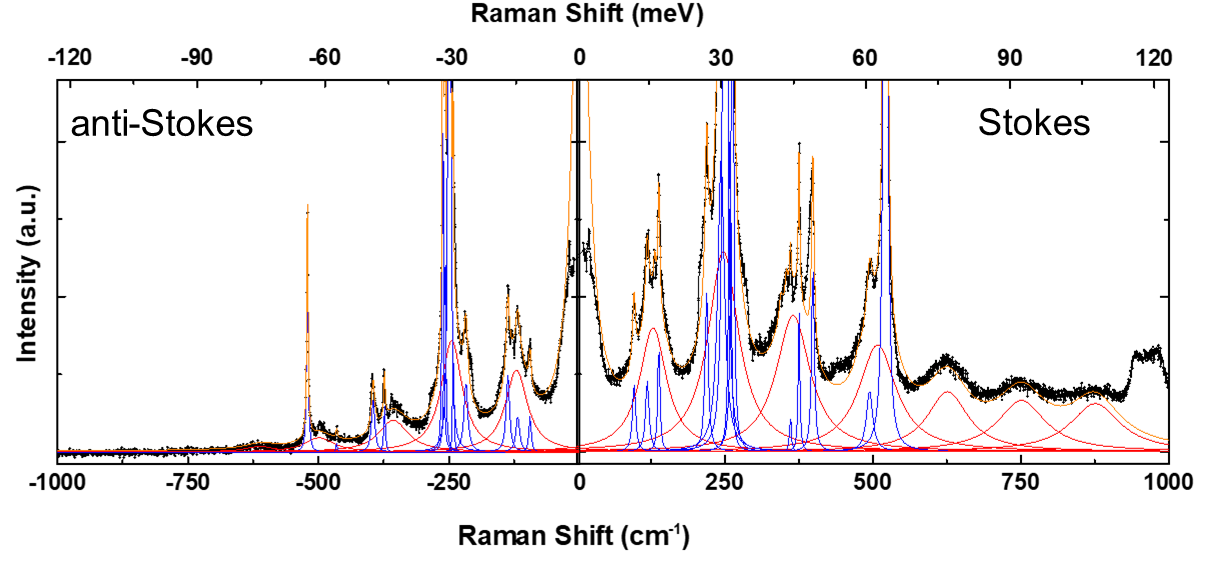}}
\caption{\label{fig:figS2} Fitted 1L-WSe$_2$/SiO$_2$/Si spectrum, collected at 295K and 532nm}
\end{figure*}
\section{Conclusions}
We investigated the light scattered and emitted 1L-WSe$_2$ excited above the free carrier gap. We detected a periodic modulation of phonon-assisted hot PL with a period$\sim$15meV both in Stokes and anti-Stokes. We measured the evolution I$_S$ and I$_{AS}$ for the periodic steps from 78 to 295K. We explained these high-order processes using a cascade model where electrons (holes) make successive transitions between real states with a finite probability of radiative recombination at each step. The electron states in the $\Lambda$ valleys play a role as intermediate states for efficient exciton relaxation. Our findings  provide fundamental understanding of the initial steps of exciton relaxation dynamics in 1L-WSe$_2$ and are valuable for tailoring optoelectronic applications based on this material.
\section{Acknowledgements} We acknowledges funding from ANR 2D-vdW-Spin, ANR MagicValley, the Institut Universitaire de France, the RFBR and CNRS joint project 20-52-16303, EU Graphene Flagship, ERC Grants Hetero2D and GSYNCOR, EPSRC Grants EP/K01711X/1, EP/ K017144/1, EP/N010345/1, EP/L016087/1.
\section{Methods}
\subsection{Raman and PL spectra fitting}
Fig.\ref{fig:figS2} shows representative data fits. The spectrum, measured at 295K at 532nm, is shown with black dots. Lorentzians are used to fit the Raman peaks (FWHM$\sim$1-10cm$^{-1}$) and are shown in blue. The residual spectral weight is also fitted with Lorentzians and results into the broader (FWHM$\sim$50-80cm$^{-1}$) peaks of the hot PL, shown in red.
\subsection{Diagrammatic calculation of Stokes scattering at 0K}
For 0K, we calculate the S emission. The light scattering cross-section can be written as (disregarding polarization dependence):
\begin{equation}
\label{scatt:gen}
\sigma(\omega_i,\omega_f)=s\sum_k S_k(\omega_i,\omega_f),
\end{equation}
where $s$ is a prefactor weakly dependent on the initial and final frequencies, $S_k(\omega_i, \omega_f)$ is the effective cross-section due to the participation of $k$ phonons in the intermediate states, shown by green arrows in Fig.\ref{fig:fig3}a. We assume that photoexcitation results in the generation of excitons due to their high (hundreds of meV) binding energies\cite{wang2018colloquium}. The description in the case of unbound e-h pairs is similar and outlined below.
\begin{figure}
\centerline{\includegraphics[width=90mm]{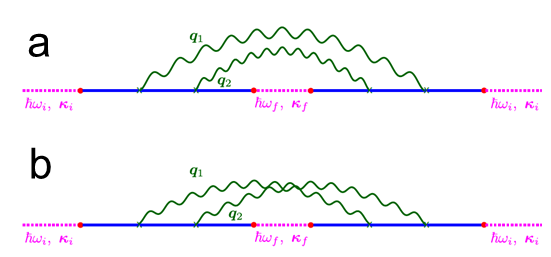}}
\caption{\label{fig:figS6} (a-b) Feynman diagrams corresponding to the two-phonon processes. The dotted magenta lines are the photon Green functions in the free space. Blue lines are exciton Green functions (retarded ones at the left-hand side, before the internal photon function, and advanced at the right-hand side). Wiggly lines are phonon Green functions $D^-$.}
\end{figure}

The calculation of the partial contributions $S_k(\omega_i,\omega_f)$ can be performed in the framework of the diagram technique of Refs.\cite{Zeyher:1975aa,ivchenko_lang_pavlov77,Goltsev_1983}. We extend their treatment to the two-dimensional case, with the exciton-phonon interaction described by the matrix element $M_0$, independent of wavevector. Phonons, as in Refs.\cite{he2020valleyphonon,PhysRevB.90.045422}, are considered as dispersionless. This is reasonable for 1L-TMDs where Fr\"ohlich coupling is suppressed\cite{7496798,PhysRevB.100.041301}. Following Ref.\cite{PhysRevB.100.041301}, we introduce the coupling constant:
\begin{equation}
\label{exc-ph}
\beta = \frac{2\mathcal S m|M_0|^2}{\hbar^3\Omega},
\end{equation}
where $\mathcal S$ is the normalization area, $m$ the exciton translational mass and $\Omega$ the phonon frequency. We focus on one excitonic band (stemming from $1s$ excitons), and disregard the multivalley structure of 1L-TMDs\cite{wang2018colloquium} for simplicity. The exciton damping rate in the state with wavevector $\bm k$ due to emission of dispersionless phonons is given by:
\begin{equation}
\begin{split}
\label{opt}
\gamma_{o,k} = \frac{1}{2\tau_{o,k}} = \frac{\pi}{\hbar} \sum_{\bm k'} |M_0|^2 \delta(E_k - E_{k'}-\hbar\Omega)=\\
\frac{\beta \Omega}{4} \Theta(E_k-\hbar\Omega),
\end{split}
\end{equation}
where $E_k=\hbar^2 k^2/2m$ is the exciton dispersion, $\Theta(x)$ is the Heaviside step function. The total damping rate of the exciton $\gamma_k>\gamma_{o,k}$ contains also the contributions due the interaction with acoustic phonons\cite{PhysRevLett.119.187402,shree2018exciton}, disorder\cite{martin2018encapsulation}, non-radiative\cite{martin2018encapsulation} and radiative (for states within the light cone) damping\cite{Schneider2018a}.

Correspondingly, the exciton retarded Green functions read:
\begin{equation}
\label{G:x}
G(\varepsilon,\bm k) = \frac{1}{\varepsilon - E_k + \mathrm i \gamma_k}.
\end{equation}
By labelling $\Gamma_q$ the phonon damping, the Green functions become:
\begin{align}
\label{D:ph}
&D(\omega,\bm q) = \frac{1}{\hbar \omega -\hbar\Omega + \mathrm i \Gamma_q}, \\
& D^-(\omega,\bm q) = D(\omega, \bm q) - D^*(\omega,\bm q) = %\\
 \frac{-2\mathrm i \Gamma_q}{(\hbar \omega -\hbar\Omega)^2+ \Gamma_q^2}.\nonumber
\end{align}
Figure~\ref{fig:figS6}a illustrates the relevant diagrams describing the two-phonon process:
\begin{subequations}
\label{S2}
\begin{multline}
\label{S2a}
S_2^{(a)} = -|M_0|^4\sum_{\bm k} \int_{-\infty}^\infty \frac{d\omega}{2\pi} G(\hbar\omega_i - E_1, \bm \kappa_i) \\
\times G(\hbar\omega_i - E_1 - \hbar \omega, \bm k)G(\hbar\omega_f- E_1, \bm \kappa_f)\\
\times D^-(\omega, \bm \kappa_i - \bm k)D^-(\omega_i - \omega_f - \omega, \bm k - \bm \kappa_f )\\
\times G^*(\hbar\omega_i - E_1, \bm \kappa_i)G^*(\hbar\omega_i - E_1 - \hbar \omega, \bm k)G^*(\hbar\omega_f- E_1, \bm \kappa_f) \\
= -|M_0|^4\left|G(\hbar\omega_i - E_1, \bm \kappa_i)\right|^2 \left|G(\hbar\omega_f- E_1, \bm \kappa_f)\right|^2 \\
\times \sum_{\bm k} \int_{-\infty}^\infty \frac{d\omega}{2\pi} D^-(\omega, \bm \kappa_i - \bm k)D^-(\omega_i - \omega_f - \omega, \bm k - \bm \kappa_f ) \\
\times \left|G(\hbar\omega_i - E_1 - \hbar \omega, \bm k)\right|^2.
\end{multline}
\begin{multline}
\label{S2b}
S_2^{(b)}
= -|M_0|^4\left|G(\hbar\omega_i - E_1, \bm \kappa_i)\right|^2 \left|G(\hbar\omega_f- E_1, \bm \kappa_f)\right|^2 \\
\times \sum_{\bm k} \int_{-\infty}^\infty \frac{d\omega}{2\pi} D^-(\omega, \bm \kappa_i - \bm k)D^-(\omega_i - \omega_f - \omega, \bm k - \bm \kappa_f ) \\
\times G(\hbar\omega_i - E_1 - \hbar \omega, \bm k)G^*(\hbar\omega_f - E_1 + \hbar \omega, \bm \kappa_i + \bm \kappa_f - \bm k)
\end{multline}
\end{subequations}
where $E_1=E_g-E_{b,1s}$ is the $1s$ exciton excitation energy, $E_{b,1s}$ is its binding energy.

Figure~\ref{fig:figS6}a describes the process where two phonons are emitted one after another. Figure~\ref{fig:figS6}b shows the quantum interference of two-photon emission processes\cite{ivchenko_lang_pavlov77}. When $\Gamma_{q}\gg\gamma_{k}$, the contribution in Fig.\ref{fig:figS6}b and given by Eq.\eqref{S2b}, is smaller by a factor $\gamma_{k}/\Gamma_{q}$, compared to the non-crossing contribution in Fig.\ref{fig:figS6}a.

We now focus on the more realistic case where $\Gamma_q\ll\gamma_k$. For simplicity we disregard the $k$-dependence of $\gamma_{k}$ and the $q$-dependence of $\Gamma_{q}$ and omit the corresponding subscripts. Thus:
\begin{subequations}
\label{S2:1}
\begin{multline}
\label{S2a:1}
S_2^{(a)} = \frac{4\Gamma}{4\Gamma^2+(2\Omega -\omega_i + \omega_f)^2} |M_0|^2\left|G(\hbar\omega_i - E_1, \bm \kappa_i)\right|^2 \\
\times \left|G(\hbar\omega_f- E_1, \bm \kappa_f)\right|^2 \\
\times \frac{1}{\hbar} \sum_{\bm k} \frac{|M_0|^2}{\hbar^2\gamma^2+(\hbar\Omega + \hbar \omega_f - E_1 - E_k)^2}
\end{multline}
\begin{multline}
\label{S2b:1}
S_2^{(a)} = \frac{4\Gamma}{4\Gamma^2+(2\Omega -\omega_i + \omega_f)^2} |M_0|^2 \\
\times \left|G(\hbar\omega_i - E_1, \bm \kappa_i)\right|^2 \left|G(\hbar\omega_f- E_1, \bm \kappa_f)\right|^2 \frac{1}{\hbar} \\
\times  \sum_{\bm k} \frac{|M_0|^2}{(\hbar\omega_f+\hbar\Omega - E_1 - E_k + \mathrm i \gamma)}\\
\times \frac{1}{(\hbar\omega_f+\hbar\Omega - E_1 - E_{k'} - \mathrm i \gamma)},
\end{multline}
\end{subequations}
with $\bm k'=\bm\kappa_i+\bm\kappa_f-\bm k$.

Note that
\begin{subequations}
\label{internal}
\begin{multline}
\frac{1}{\hbar} \sum_{\bm k} \frac{|M_0|^2}{\hbar^2\gamma^2+(\hbar\Omega + \hbar \omega_f - E_1 - E_k)^2} = \frac{\gamma_o}{\gamma} , \\
\frac{1}{\hbar} \sum_{\bm k} \frac{|M_0|^2}{(\hbar\omega_f+\hbar\Omega - E_1 - E_k + \mathrm i \gamma)}\\
\times \frac{1}{(\hbar\omega_f+\hbar\Omega - E_1 - E_{k'} - \mathrm i \gamma)} = \\
\frac{\gamma_o/\gamma}{\sqrt{1+\left(\frac{|\bm \kappa_i + \bm \kappa_f|v}{2\gamma} \right)^2}},
\end{multline}
\end{subequations}
where $v = \sqrt{2(\hbar\omega_f + \hbar \Omega - E_1)/m}$.
Thus:
\begin{subequations}
\label{S2:2}
\begin{equation}
\begin{split}
\label{S2a:2}
S_2^{(a)}=\frac{4\Gamma}{4\Gamma^2+(2\Omega -\omega_i + \omega_f)^2} |M_0|^2 \\
\times \frac{\gamma_o}{\gamma}\left|G(\hbar\omega_i - E_1, \bm \kappa_i)\right|^2 \left|G(\hbar\omega_f- E_1, \bm \kappa_f)\right|^2,
\end{split}	
\end{equation}
\begin{equation}
\label{S2b:2}
S_2^{(b)}=\frac{S_2^{(a)}}{\sqrt{1+\left(\frac{|\bm \kappa_i + \bm \kappa_f|v}{2\gamma} \right)^2}}.
\end{equation}
\end{subequations}
The factor
\[
\frac{|\bm \kappa_i + \bm \kappa_f|v}{2\gamma} \sim \frac{l_{\rm eff}}{\lambda},
\]
where $l_{\rm eff}$ is the effective mean free path of the exciton, and $\lambda$ is the characteristic wavelength of light. In backscattering in plane, since scattered light is emitted by both 1L-WSe$_2$ sides:
\begin{equation}
\label{back}
\bm \kappa_f=-\bm\kappa_i,
\end{equation}
the diagram with crossed phonon lines doubles the result stemming from the diagram in Fig.\ref{fig:figS6}b. This is due to coherent backscattering (or weak localization) effect\cite{1977ZhETF..72.2230I,glazov2020quantum}. Otherwise the contribution of the diagram Fig.\ref{fig:figS6}b is negligible, provided that $l_{\rm eff}\gg\lambda$. While the latter condition may not be strictly fulfilled in 1L-TMDs\cite{glazov2020quantum}, we disregard the contributions due to the crossed diagrams to provide an analytical model.

Importantly, the phonon scattering is resonant, taking place via real intermediate states and, accordingly, the scattering cross-section acquires a factor $\frac{\gamma_o}{\gamma}$, which gives the probability for an exciton to emit a phonon during its lifetime in a state with wavevector $\bm k$. If inelastic scattering is dominated by a single phonon mode, $\gamma_o/\gamma$ can be close to unity.
\begin{figure}
\centerline{\includegraphics[width=90mm]{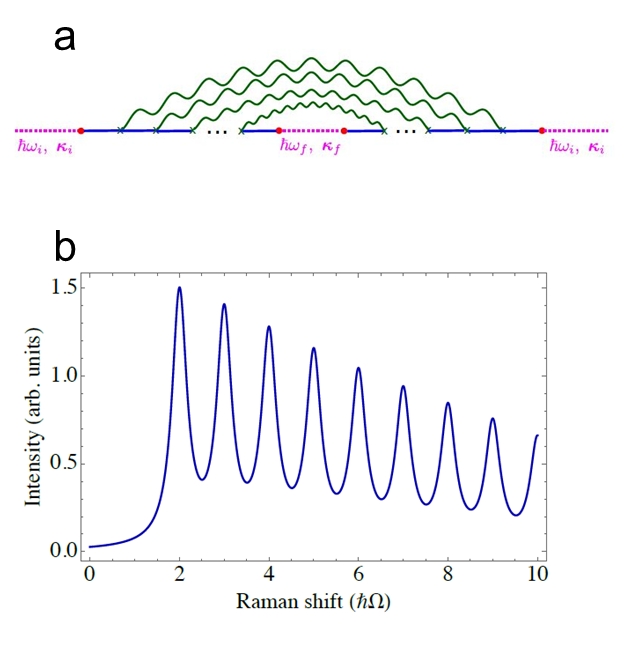}}
\caption{\label{fig:figS7}(a) Multiphonon scattering in the non-crossing approximation. (b) Hot PL for $\gamma_o/\gamma=0.9$, $\Gamma/\Omega=0.1$, $K=10$.}
\end{figure}

We now consider multiphonon processes. Within the non-crossing approximation, where we take into account the diagrams where the phonon propagators do not cross, i.e., we disregard the interference of the phonons, we can sum the contributions of the diagrams in Fig.\ref{fig:figS7}a with $k=2,3,\ldots$ phonon lines.

The maximum number of phonons involved in the process is given by Eq.~\eqref{steps:K}.
%:
%\begin{equation}
%\label{steps:K}
%K=\left\lfloor \frac{\hbar\omega_i - E_1}{\hbar\Omega}\right\rfloor.
%\end{equation}
Performing the calculations analogous to those presented above we get Eq.~\eqref{sigma}. In particular
%:
%\begin{equation}
%\begin{split}
%\label{sigma}
%\sigma(\omega_i,\omega_f) = \\
%\sigma_0(\omega_i,\omega_f) \sum_{k=2}^K \frac{1}{\pi} \frac{2\Gamma}{4\Gamma^2+(k\Omega -\omega_i + \omega_f)^2}\left(\frac{\gamma_o}{\gamma}\right)^{k-1},
%\end{split}
%\end{equation}
%where $\sigma_0(\omega_i,\omega_f)$ is a smooth function of frequency.
%
%
for $K\to \infty$ we get:
\begin{equation}
\begin{split}
\label{sigma:asympt}
\sigma(\omega_i,\omega_f) =\sigma_0(\omega_i,\omega_f) \frac{x}{\pi} \\
\times \Re\left\{\frac{ _2F_1\left(1,\frac{-2\mathrm i \Gamma + 2\Omega - \omega_i + \omega_f}{\Omega},\frac{-2\mathrm i \Gamma + 3\Omega - \omega_i + \omega_f}{\Omega},x\right)}{2\Gamma +\mathrm i (2\Omega - \omega_i + \omega_f)}\right\} ,
\end{split}
\end{equation}
with $_2F_1(a,b,c,x)$ the hypergeometric function, and $x=\gamma_o/\gamma$.

A typical calculated spectrum at 0K considering the S component of the emission is in Fig.\ref{fig:figS7}b. Each cascade step provides a factor $\gamma_o/\gamma$ to the scattering cross-section. If the scattering rates $\gamma_o$ and $\gamma$ are energy dependent, the S component of emission at $j$th step is given by the products 
\begin{equation}
\label{S:gamma:n}
\mathrm I_j \propto \prod_{k=2}^j \frac{\gamma_o(k)}{\gamma(k)},
\end{equation}
where the argument $k$ denotes the step of the cascade (i.e., the energy) where the corresponding scattering rate is taken.
 
\begin{figure*}
\centerline{\includegraphics[width=160mm]{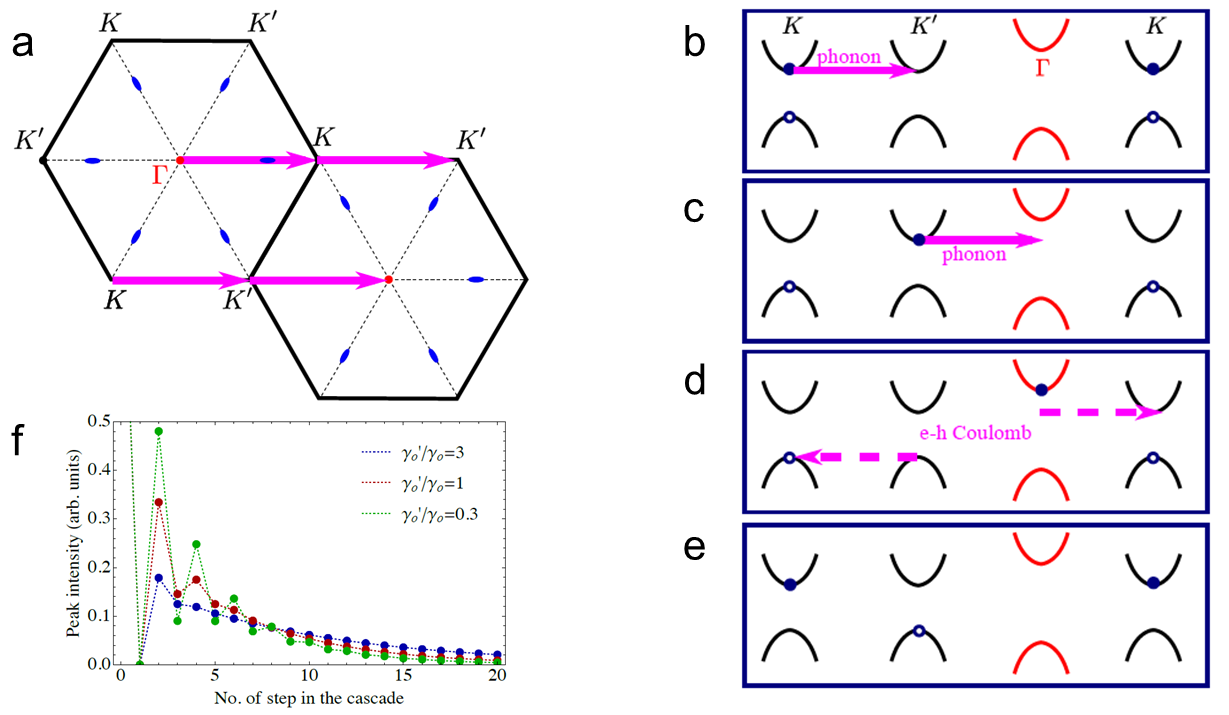}}
\caption{\label{fig:figS9} Schematic band diagram and scattering pathways in e-h representation. (a) Extended BZ with $K$ valley shown twice for convenience. (b) Initial state $\Gamma$ ($K-K$) exciton. (c) After first phonon scattering $K$ ($K'-K$) exciton. (d) After second phonon scattering we have $K'$-like ($\Gamma-K$) exciton. (e) e-h interaction brings a $\Gamma-K$ exciton to the $K'$-exciton ($K-K'$). Open circle denotes empty state (h state can be obtained by time-reversal). (f) Scattered I accounting for $\Gamma\to K\to K'$ photon-assisted transitions for $\gamma'/\gamma_o=0.3$, $\gamma''/\gamma_o=0.3$, at 0K}
\end{figure*}

In the presence of static disorder, or quasi-elastic acoustic phonon scattering with negligible energy transfer, additional diagrams with the corresponding scattering processes should be taken into account. To illustrate that elastic scattering processes do not suppress oscillations in the Raman and hot PL, we consider exciton scattering by static impurities with scattering rate $\gamma_{\rm imp}$, with $\gamma$, the total exciton scattering rate. Taking into account diagrams similar to Fig.\ref{fig:figS7}a, but with impurity lines, $\gamma_o/\gamma$ is renormalized by elastic scattering as:
\begin{equation}
\label{rat:renorm}
\frac{\gamma_o}{\gamma} \to \frac{\gamma_o}{\gamma} \sum_{n=0}^\infty \left(\frac{\gamma_{\rm imp}} {\gamma} \right)^n = \frac{\gamma_o}{\gamma - \gamma_{\rm imp}}.
\end{equation}
This means that for elastic scattering the exciton energy does not change, thus it does not smear-out the I oscillations.
\subsection{Kinetic equation}
The non-crossing approximation corresponds to a kinetic equation model where the exciton dynamics after optical excitation is described by the Boltzmann equation in the form\cite{tserkovnikov1993kinetic}:
\begin{equation}
\label{kinetic:gen}
\gamma' f(\bm k) = \sum_{\bm k'} W_{\bm k\bm k'} [f_{\bm k'} - W_{\bm k'\bm k} f_{\bm k}] + g_{\bm k}.
\end{equation}
with $W_{\bm k'\bm k}$ the transition rate between states with wavevectors $\bm k$ and $\bm k'$, with phonon emission or absorption. For simplicity, we disregard elastic scattering. $\gamma'$ is the damping rate unrelated to exciton-phonon interaction, $g_{\bm k}$ is the exciton generation rate. The kinetic equation also allows us to account for finite T effects\cite{tserkovnikov1993kinetic}.

It is convenient to average $f_{\bm k}$ over the in-plane directions of $\bm k$, and consider just the exciton energy distribution function $f(\varepsilon)$. The latter can be recast as [cf. Eq.~\eqref{distribution}]:
\begin{equation}
\label{distribution:SI}
f(\varepsilon)=\sum_{j=-\infty}^\infty f_j \delta(\varepsilon_0- j\hbar\Omega),
\end{equation}
with $f_j$ satisfying the set of equations:
\begin{equation}
\begin{split}
\label{fj:eqs}
\gamma f_j = \gamma_o \left[f_{j-1} (n_o+1) + f_{j+1} n_o\right] + g_j, \\
\quad j=\ldots, -2,-1,0,1,2,\ldots,
\end{split}
\end{equation}
where\[n_o=\frac{1}{\exp{\left(\frac{\hbar\Omega}{k_B T}\right)}-1},\] is the phonon mode occupancy at T, $\gamma_o$ is the rate of spontaneous phonon emission,\[\gamma=\gamma_o (2n_o+1)+\gamma',\] is the total exciton damping rate. $g_j$ is the exciton generation rate in the state $j$ related to the process of virtual formation of the exciton within the light cone, and its consequent relaxation to the real state with the phonon emission or absorption. Thus, the non-zero values of $g_j$ are:
\begin{equation}
\label{generation}
g_1 = g (n_o+1), \quad g_{-1} = g n_o,
\end{equation}
with $g$ a parameter. In the main text we replaced the generation rate Eq.\eqref{generation} with a simplified model with $g_0=g \ne 0$. This is also valid if elastic scattering is strong and excitons can leave the light cone via static defect scattering.

Since photon emission requires a phonon-induced transition, I of peak $j$ with phonon-induced energy shift $j\hbar\Omega$ is given by:
\begin{equation}
I_j = f_{j-1} \gamma_o(n_o+1) + f_{j+1} \gamma_o n_o \propto f_j.
\end{equation}
The last proportionality is due to the kinetic Eq.\eqref{fj:eqs} and is valid for $j \ne 0, \pm 1$.

We now address I of $|j|=1$ peaks. We consider $I_{j=1}/I_{j=2}$ as plotted in Fig.\ref{fig:figS8}. The mechanisms of $j=1$ peak formation are as follows. (i) Elastic disorder-induced scattering, Eq.\eqref{rat:renorm}, which provides a transfer between states within the light cone and states at the dispersion. (ii) Combination of phonon emission and absorption, where the $j=1$ peak appears as a result of two phonon emission followed by one phonon absorption. In (i) $I_{j=1}/I_{j=2}$ is not dependent on T, while in (ii):
\begin{equation}
\label{rat12}
\frac{I_{j=1}}{I_{j=2}} = \exp{\left(-\frac{\hbar\Omega}{k_B T}\right)},
\end{equation}
strongly depends on T. The results of Eq.\eqref{rat12} and plotted in Fig.\ref{fig:figS8} by a solid line and agree with experiments. Elastic processes could be the origin of a small offset between the experiment and the fitted curve.
\subsection{Cascades in a free carriers model}\label{SI:theory:unbound}
Cascades in phonon-assisted hot PL are possible for free carriers, i.e., for e-h pairs in the continuum states. In this situation the phonon-assisted e/h relaxation is independent, and governed by Eq.\eqref{fj:eqs}. Light scattering can be represented as
\[\mbox{photon} \to \mbox{electron,} \bm k + \mbox{hole,} \bm k\]
\[
 \xrightarrow{\hbar\Omega} \mbox{electron,} \bm k-\bm q + \mbox{hole,} \bm k \ldots\]\[\xrightarrow{\hbar\Omega} \mbox{electron,} \bm k + \mbox{hole,} \bm k\to \mbox{photon}.\] 
Accordingly, we expect oscillations in scattered light I with period $\hbar\Omega$.
\subsection{Intervalley scattering model}\label{SI:theory:valley}
We now consider the case of exciton scattering enabled by BZ edge phonons with wavevectors $\bm Q\sim\pm\bm K$. An exciton scatters between the $\Gamma$ valley ($K-K$ or $K'-K'$ exciton) and $K/K'$ valleys ($K'-K$ and $K-K'$ excitons), Fig.\ref{fig:figS9}a. Photon emission occurs only for the states with small wavevectors ($\sim$the photon ones), according to the paths:
\[\mbox{photon} \to \Gamma \xrightarrow{\hbar\Omega} K \xrightarrow{\hbar\Omega} \mbox{photon};\]
\[\mbox{photon} \to \Gamma \xrightarrow{\hbar\Omega} K \xrightarrow{\hbar\Omega} K' \xrightarrow{\hbar\Omega} \mbox{photon};\]
\[\mbox{photon} \to \Gamma \xrightarrow{\hbar\Omega} K \xrightarrow{\hbar\Omega} \Gamma \xrightarrow{\hbar\Omega} K \xrightarrow{\hbar\Omega} \mbox{photon};\]
\[\mbox{photon} \to \Gamma \xrightarrow{\hbar\Omega} K \xrightarrow{\hbar\Omega} \Gamma \xrightarrow{\hbar\Omega} K' \xrightarrow{\hbar\Omega} \mbox{photon}; \ldots\]
Processes like ``$\Gamma \xrightarrow{\hbar\Omega} \mbox{photon}$'' are forbidden due to momentum conservation.

The set of equations describing the processes is:
\begin{subequations}
\label{set:G:K:K'}
\begin{multline}
\gamma_\Gamma f_j^\Gamma =\gamma_o \left[f_{j-1}^{K} (n_o+1) + f_{j+1}^{K} n_o\right] +\\ \gamma_o \left[f_{j-1}^{K'} (n_o+1) + f_{j+1}^{K'} n_o\right] + g\delta_{j,0},\\
\gamma_{K} f_j^{K} =\gamma_o \left[f_{j-1}^{\Gamma} (n_o+1) + f_{j+1}^{\Gamma} n_o\right] +\\ \gamma_o' \left[f_{j-1}^{K'} (n_o+1) + f_{j+1}^{K'} n_o\right],\\
\gamma_{K'} f_j^{K'} =\gamma_o \left[f_{j-1}^{\Gamma} (n_o+1) + f_{j+1}^{\Gamma} n_o\right] +\\ \gamma_o' \left[f_{j-1}^{K} (n_o+1) + f_{j+1}^{K} n_o\right].
\end{multline}
\end{subequations}
with $f_j^\Gamma$, $f_j^K$ and $f_j^{K'}$ the exciton occupancies in the corresponding valley. $\gamma_o$ and $\gamma_o'$ are the photon spontaneous emission rates for the scattering processes $\Gamma \leftrightarrow K/K'$ and $K \leftrightarrow K'$, respectively, and the decay rates\[\gamma_\Gamma=\gamma_o (2n_o+1)+\gamma', \quad \gamma_{K}=\gamma_{K'} = (\gamma_o +\gamma_o')(2n_o+1) +\gamma''.\]

The scattering processes in the e-h picture are presented in Figs.\ref{fig:figS9}b-e. These demonstrate that $\gamma_o$ and $\gamma_o'$ are different because the second process ($K\leftrightarrow K'$) needs additional Coulomb interaction, Figs.\ref{fig:figS9}c-e. Thus, direct phonon-induced $K\leftrightarrow K'$ transfer is impossible, and the corresponding process takes place via an intermediate state with e in the $\Gamma$ valley. The results of the calculations in Fig.\ref{fig:figS9}f demonstrate that pronounced I oscillations should take place at low T (below the phonon energy). These oscillations are not observed experimentally in Fig.\ref{fig:fig2}, ruling-out this pathway.

%\bibliography{casc_bib}  


\end{document}